\documentclass[aps,onecolumn,superscriptaddress,amsmath,showpacs,tightenlines]{revtex4-1}
\usepackage{amssymb}
\usepackage{amsmath}
\usepackage{graphicx}
\usepackage{subfigure}
\usepackage{natbib}
\usepackage{epsfig}
\usepackage{amsfonts}
\usepackage{mathrsfs}
\usepackage{epstopdf}
\usepackage{xcolor}
\usepackage[toc,page,title,titletoc,header]{appendix}

\begin{document}

 \title{Quantifying the potential and flux landscapes for nonequilibrium multiverse, a new scenario for time arrow}
\author{Hong \surname {Wang}}
\affiliation{State Key Laboratory of Electroanalytical Chemistry, Changchun Institute of Applied Chemistry, Chinese Academy of Sciences, Changchun 130022, China}
\author{Xinyu \surname{Li}}
\affiliation{College of Physics, Jilin University, 130012,China}
\author{Jin \surname{Wang}}
\email{Email: jin.wang.1@stonybrook.edu}
\affiliation{Department of Chemistry and Department of Physics and Astronomy, State University of New York at Stony Brook, NY 11794, USA}

\begin{abstract}
We propose a new scenario of nonequilibirum multiverse. We quantified the potential landscape and the flux landscape for the multiverse. The potential landscape quantifies the weight of each universe. When the terminal vacuum with zero (flat) or negative cosmological constant (AdS) have a chance to tunnel back to the normal universes with positive cosmological constant (dS) through the bounce suggested by the recent studies, the detailed balance of the populations of the multiverse can be broken. We found that the degree of the detailed balance breaking can be quantified by the underlying average flux and associated flux landscape, which gives arise to the dynamical origin of irreversibility and the time arrow of the multiverse. We also showed that the steady state of the multiverse is maintained by the thermodynamic cost quantified by the entropy production rate which is associated to the flux. This gives arise to thermodynamic origin of time irreversibility. On the other hand, we show that the evolution dynamics of the multiverse is determined by both the potential landscape and flux landscape. While the potential landscape determines the weight of the universes in the multiverse and attracts the multiverse to the steady state basins, the flux landscape provides the cycles or loops associating certain universes together. We show that terminal vacuum universes can have dominant weights or lowest potentials giving arise to a funnel shaped potential landscape, while terminal vacuum universes together with other normal universes including ours can form dominant cycles giving arise to a funnel shaped cycle flux landscape. This indicates that even our universe may not be distinct from others based on the probability measure, it may lie in the dominant cycle(s), leading to higher chance of being found. This may provide an additional way beyond the anthropic principle for identifying our universe.
\end{abstract}

\maketitle

%\tableofcontents

\section{Introduction}

The conventional Big Bang cosmology theory faces difficulty for interpreting the homogeneity problem, the flatness problem, and the magnetic monopole problem and so on. Inflation theory was suggested to resolve these issues~\cite{AH}. The essential picture of the inflation is that the universe went through a transient fast accelerating expansion phase soon after the Big Bang. The inflation is often thought as being driven by the vacuum energy/dark energy. If there are different vacuum states, different vacuum can tunnel to each other~\cite{SR}. Following the inflation of the universe, the process of vacuum decay is reminiscent to the bubble formation in vaporization of liquid to gas phase transitions. However, there are issues associated to this old inflation model.

Although the false vacuum can decay to the true vacuum through the bubble nucleation, They cannot have the chance to collide and reheat the universe since they cannot catch up with the expansion of the rest of the inflationary universe. For this, a new inflation theory was proposed to have a slow and continuous transition from the false vacuum to the true vacuum within a single bubble (universe)~\cite{AD}. The inflation theory predicted the amplitudes and the fluctuation spectrum of the cosmic background radiation as the seeds for large structure formation in our universe consistent with the observations.

Furthermore, it was pointed out that the inflation can continue without ending in most part of the universe, giving arise to new bubble universes one after another. Thus, the inflation theory gives rise to a very different picture of the evolution of the universe, leading to a multiverse with eternal inflation~\cite{AD,SW}.

On the other hand, in string theory, there are huge number of different kinds of universes with different vacuum and different coupling constants. In fact, there are estimated of $10^{500}$ of such universes or vacuum. Some studies suggest that there is no preferred universe and all universes should be treated on an equal footing~\cite{SW,AV}.

There are immediate questions about how our universe can be identified from the enormous amount of the possible other universes~\cite{SW}. There is another closely related issue. According to the quantum field theory, the vacuum energy is huge. But according to the cosmological observation, the cosmological constant is tiny. The associated issue is why the cosmological constant of our universe is so small out of so many possibilities of the different universes~\cite{SW}. Anthropic principle has been used to explain the observational existence of our universe~\cite{SW}. For example, if we have somewhat different universe with different cosmological constant and coupling constants, the galaxies will not be formed properly, our human being will not come to existence to be able to observe our current universe. Therefore, according to the anthropic principle, we happen to live in a universe which can produce ourselves and the observational universe we currently see. This has been used to explain why the cosmological constant of our universe is so small~\cite{SW}.

Despite of progresses being made, there are still challenging issues related to the multiverse picture. One is related to the observational evidences. Suggestions have been made to study the emergence of the black holes as the trace of the possible consequence of the collisions of different universes~\cite{DP}. The clear evidences remain to be seen.

Another important issue is related to the time arrow. How can we determine the direction of time in the multiverse? Suggestions have been made to explain the time arrow by including the terminal universes with negative or zero cosmological constants~\cite{LS}. These terminal universes cannot convert to the universes through the vacuum tunneling. But the other universes can tunnel to these terminal universes~\cite{LS}. Therefore, the terminal universes act like sinks in the multiverse. This produces the irreversibility and time arrow of the multiverse. However, the time arrow can only last at finite time during the multiverse evolution. In the long time limit, the time arrow ceases to exist.

Recent studies suggested the possibilities of bounces avoiding singularities from a contracting universe (with negative cosmological constant) back to the expanding universe (with positive cosmological constant)~\cite{JG}. Furthermore, there were also discussions on the possibility of expanding universe (with positive cosmological constant) born from the flat universe (with zero cosmological constant)~\cite{EF,EFA,WF,WFI,AL,FD,SFVF}. If one takes this into consideration, the terminal vacuum (with zero or negative cosmological constant) will have small chances of coming back to the normal vacuum (with positive cosmological constant). The detailed balance can be broken in this case~\cite{JG,FD,SFVF}. This suggests a need to study the multiverse problem from a nonequilibrium perspective.

In this work, we will study the multiverse evolution from the perspective of nonequilibrium physics. Without the assumption of the detailed balance among the vacuum tunneling switching of the universes, we show that the multiverse evolution is driven by two forces. One force is the underlying probability landscape of the multiverse quantified by the steady state probability distribution in multiverse state space. The other is the steady state probability flux which quantifies the degree of the detailed balance breaking. While the landscape attracts the multiverse down to their steady state basins of attractions, the flux provides a driving force for the cycle flow in multiverse state space. The steady state probability forms a potential or weight landscape while the steady state flux forms a flux landscape of cycles. We found that while the terminal vacuum has largest weight and therefore lowest potential, the largest flux cycles can include not only terminal vacuum but also normal vacuum with positive cosmological constant including our universe. Therefore, although our universe might not be the highest probability one, it may lie in the highest probable flux cycle, which still gives a higher chance to be observed. On the other hand, the explicit detailed balance breaking characterized by the magnitude of the flux through the possibility of terminal vacuum tunneling back to normal universe provides a source of irreversibility and therefore the arrow of time at all times (even at long time limit). Moreover, the dynamical driving force in terms of flux gives arise to the entropy production rate. In fact, the entropy production rate provides the thermodynamic cost for maintaining the steady state of the multiverse, which is also an indication of the time arrow of the multiverse.

\section{The Model}
%\label{sec:1}
%and \cite{RefJ}
\subsection{Multiverse description}
\begin{figure}[tbp]
\centering
\includegraphics[width=7cm]{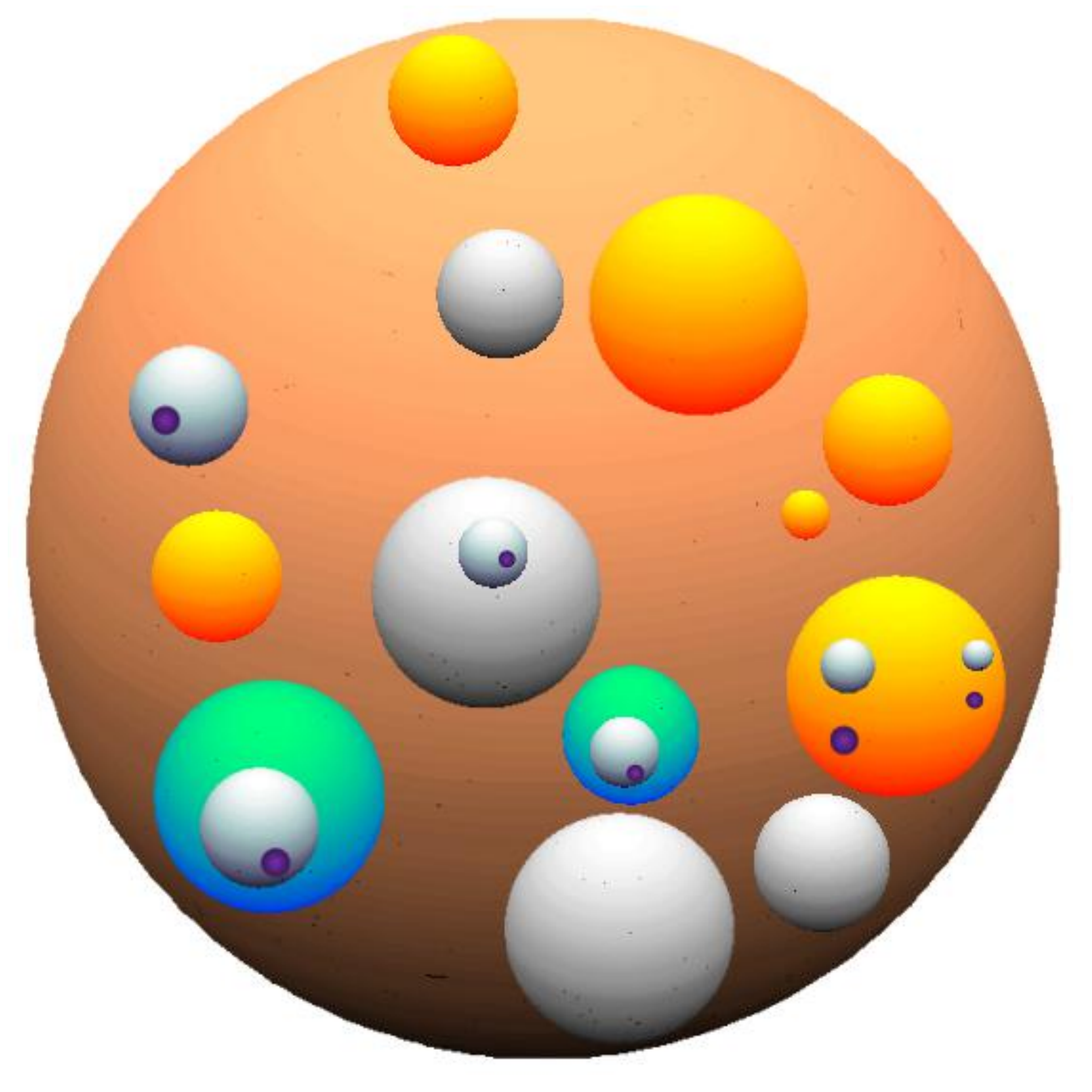}
\caption{\label{fig:i} The multiverse picture of the bubble universes.}
\end{figure}

%\label{sec:2}
%as required. Don't forget to give each section
%and subsection a unique label (see Sect.~\ref{sec:1}).
%To explore the multiverse evolution, let us first consider the evolution of the universe. In general, we can use Einstein's general relativity to describe the evolution of universe. Let us assume that the universe is homogeneous and isotropic on the large scales and treat matter as the perfect fluid~\cite{UE}. From the Einstein's equations, we can obtain the Friedmann Equations (flat universe) for the evolution of the universe~\cite{UE}:

\begin{eqnarray}
\label{eq:2.1}
 H^2=(\frac{\dot{a}}{a})^2=\frac{1}{3}\rho,
\end{eqnarray}
\begin{eqnarray}
  \dot{H}+H^2=\frac{\ddot{a}}{a}=-\frac{1}{6}(\rho+3P),
\end{eqnarray}
where \textit{a} is the expansion factor and \emph{H} is the Hubble constant representing the rate of expansion, $\rho$ is the proper energy density and \emph{P} is the pressure. If we assume that the evolution of the universe has an inflation driven by a constant vacuum energy ($\Lambda$), then \emph{H}=$\sqrt{\Lambda/3}$, \textit{a}(\textit{t})$\propto$\textit{exp}(\emph{H}t). The universe rapidly expands. Suppose there are two adjacent places, Inflation will drive them away from each other.  Beyond \emph{H}$^{-1}$, there is no causal relation~\cite{AD}. Within the universe (\emph{H}$^{-1}$), there are chances of the birth of another universe through quantum tunneling such as the ones illustrated from brown to yellow in figure~\ref{fig:i}~\cite{DP}. The yellow universe also grows rapidly, but it never catches up with the brown universe. This process continues with the birth of more other universes represented by different colors shown in figure~\ref{fig:i}~\cite{DP}. As a result, this eternal inflation leads to the multiverse landscape, a foam of expanding bubble universes within bubble universes.

 It was suggested that the cosmological constant $\Lambda$  has a discrete spectrum based on the string theory~\cite{RB},
\begin{eqnarray}
  \Lambda=\Lambda_{bare}+\frac{1}{2}\sum_{m=1}^Jn_{m}^2q_{m}^2.
\end{eqnarray}
Here $\Lambda$$_{\textit{bare}}$ is the bare cosmological constant.  \textit{q}$_{m}$ is the charge of the string compactified vacuum flux. The range of \textit{n}$_{m}$ is an integer assumed to be within
\begin{eqnarray}
 -N<n_{m}<N,
\end{eqnarray}
\emph{N} is a positive integer. The state characterized by the cosmological constant $\Lambda_{i}$ has a degree of degeneracy \emph{D}(\emph{i})~\cite{DS}:
\begin{eqnarray}
  D(i)=2^{d(i)},
\end{eqnarray}
\begin{eqnarray}
  d(i)=J-\sum_{m=1}^J\delta_{0n_{m}}.
\end{eqnarray}

We often view the cosmological constant as the vacuum energy. Different cosmological constants $\Lambda$$_{i}$ correspond to different vacuum states. These different vacua can be classified into three categories. The vacuum with positive cosmological constant $\Lambda$$_{i}$$>$0 is called the de-Sitter (dS) vacuum. The vacuum with zero cosmological constant $\Lambda$$_{i}$=0 is called the Minkowski vacuum. The vacuum with negative cosmological constant $\Lambda$$_{i}$$<$0 is called the anti-de-Sitter (AdS) vacuum. In this study, we use the Greek letters to label the dS vacuum and Latin letters to label any vacuum state.

\subsection{The probabilistic evolution of multiverse dictated by the master equation of the vacuum dynamics}

Based on the formula (3), there is an intrinsic distribution of the vacuum state. There are different definitions to describe the fractions of different vacua~\cite{MC}. For simplicity, we consider the fraction of comoving volume (\emph{P}$_{i}$). For the description by the fraction of proper volume, see~\cite{MC,JL}. The probabilistic evolution of the multiverse containing various vacua for dS vacua is determined by the master equation~\cite{FD,DS,JGA}

\begin{eqnarray}
  \frac{dP_{\alpha}}{dt}=\sum_{\beta}\kappa_{\alpha\beta}P_{\beta}-\sum_{\beta}\kappa_{\beta\alpha}P_{\alpha},
\end{eqnarray}
where $\kappa_{\alpha\beta}$ is the transition rate or the transition probability per unit time for an observer who is currently in vacuum $\beta$ to find himself in vacuum $\alpha$, it is determined by~\cite{JG,FD,DS,JGA}
\begin{eqnarray}
  \kappa_{\alpha\beta}=\frac{4\pi}{3}H_{\beta}^{-3}\Gamma_{\alpha\beta},
\end{eqnarray}
and $\Gamma_{\alpha\beta}$ is the tunneling transition rate per unit physical space time volume between two different dS vacuum, it is determined by~\cite{SR,JG,DS,ALI}
\begin{eqnarray}
\Gamma_{\alpha\beta}={\rm A_{\alpha\beta}}\cdot e^{-{\rm I_{\alpha\beta}}+S_{\beta}}.
\end{eqnarray}
Here ${\rm A_{\alpha\beta}}$ is a prefactor, ${\rm I_{\alpha\beta}}$ is the action of Coleman-DeLuccia instanton. The Coleman-DeLuccia instanton action ${\rm I_{\alpha\beta}}$ and the prefactor are the same when we interchange $\alpha$ and $\beta$~\cite{KW}, ${\rm A_{\alpha\beta}=A_{\beta\alpha},\quad   I_{\alpha\beta}=I_{\beta\alpha}} $. This is because the same bounce controls forward and backward instanton nucleations as well as their corresponding first order quantum corrections.  ${\rm S_{\beta}}$ is the Euclidean action of the dS vacuum with the  cosmological constant $\Lambda_{\beta}$.
\begin{eqnarray}
S_{\beta}=\int dx^{4}\sqrt{|g|}\cdot (-\Lambda_{\beta}) .
\end{eqnarray}
4-dimensional Euclidean dS universe could be viewed as a 5-dimensional sphere's surface in higher dimensional Euclidean space,
\begin{eqnarray}
 x_{1}^{2}+x_{2}^{2}+x_{3}^{2}+x_{4}^{2}+x_{5}^{2}=\xi^{2} .
\end{eqnarray}
$\xi$ is the radius of the 5-dimensional Euclidean sphere. According to the Hubble law, $\xi=H^{-1}$.  The volume of the 4-dimensional dS universe is the surface area of the 5-dimensional Euclidean sphere,
\begin{eqnarray}
 \int dx^{4}\sqrt{|g|}=\frac{2\pi^{\frac{4+1}{2}}}{\Gamma(\frac{4+1}{2})}\cdot H^{-4}.
\end{eqnarray}
Combine with the Friedmann equation ~\ref{eq:2.1}, and set $\rho=\Lambda_{\beta}$, we have
\begin{eqnarray}
S_{\beta}=-\frac{8\pi^{2}}{H_{\beta}^{2}}.
\end{eqnarray}

The physical meaning of the equation (7) is clear. The change in the chance or the probability of the universe $\alpha$ being observed will be determined by the input from the chances of other universes decaying to the current universe ($\alpha$) subtracting the output from decaying of the current universe ($\alpha$) to the other universes. Considering the state with cosmological constant $\Lambda_{\alpha}$ under the degree of degeneracy of $D(\alpha)$, the equation (8) should be modified as
\begin{eqnarray}
  \kappa_{\alpha\beta}=\frac{4\pi}{3}H_{\beta}^{-3}\Gamma_{\alpha\beta}\cdot D(\alpha).
\end{eqnarray}
One can easily show that
\begin{eqnarray}
  P_{\alpha}\propto H_{\alpha}^{3}exp\{\frac{24\pi^{2}}{\Lambda_{\alpha}}\}\cdot D(\alpha)
\end{eqnarray}
is the steady state solution of the master equation (7) and
\begin{eqnarray}
\frac{\kappa_{\alpha\beta}}{\kappa_{\beta\alpha}}=\frac{P_{\alpha}}{P_{\beta}}.
\end{eqnarray}
From equation (15), we know that the factor $\emph{\emph{I}}_{\alpha\beta}$ in $\Gamma_{\alpha\beta}$ does not influence $P_{\alpha}$. Therefore, we can neglect it. In addition, we know that $S_{\beta}=-8\pi^{2}/H_{\beta}^{2}$ and $H_{\beta}=\sqrt{\Lambda_{\beta}/3}$~\cite{DS}, so we can simply set
\begin{eqnarray}
\Gamma_{\alpha\beta}=exp\{S_{\beta}\}=exp\{-\frac{24\pi^{2}}{\Lambda_{\beta}}\}.
\end{eqnarray}

If all the universes are dS vacuum, the detailed balance is preserved. The resulting steady state is an equilibrium state. Since this is an equilibrium state, there is no emergence of time arrow in the multiverse. To resolve this issue, one can explore the effects of the emergence of the Minkowski vacuum or AdS vacuum.

The general master equation for the probability evolution of the multiverse involving all possible universes (dS, AdS and Minkowski) is given by
\begin{eqnarray}
 \frac{dP_{i}}{dt}=\sum_{j}\kappa_{ij}P_{j}-\sum_{j}\kappa_{ji}P_{i}.
\end{eqnarray}
In previous studies~\cite{LS,DS,JGA}, the tunneling transition rates from Minkowski and AdS universes to dS universes are assumed to be zero. Recent studies show that there is a possibility of the tunneling transition from Minkowski or AdS vacuum to dS universes through the bounce~\cite{JG,EF,EFA,WF,WFI,AL,FD,SFVF}. In~\cite{SFVF}, the authors showed that if we choose suitable boundary conditions, the detailed balance can be broken down. In this study, for convenience, we simply assume that:
\begin{eqnarray}
\kappa_{11}=S_{1}, \quad  \kappa_{i1}=S_{2},  \quad \kappa_{12}=S_{3}, \quad \kappa_{i2}=S_{4}.
\end{eqnarray}
In equation (19), state ``1" represents the AdS vacuum state. State ``2" represents the Minkowski vacuum state. The Latin letters ``$i$" represent any vacuum state. $S_{1,2,3,4}$ are certain small constants. Noted that although equation (19) appears to be a numerical setting, it contains the main point that the tunneling may occur from Minkowski (or AdS) universes to dS universes which  can break down the detail balance~\cite{SFVF} under proper boundary conditions. This feature is captured in the currernt work. In addition, we assume that the formula (17) can be generalized to
\begin{eqnarray}
\Gamma_{i\beta}=exp\{S_{\beta}\}=exp\{-\frac{24\pi^{2}}{\Lambda_{\beta}}\}.
\end{eqnarray}

\section{Results}
\subsection{Potential landscape and flux as the driving forces for the evolution of multiverse}

\subsubsection{Potential landscape and flux decomposition for the driving force dictating the evolution of multiverse.}

We can write the master equation in (18) for determining the probabilistic evolution of the multiverse involving various vacua in the form of~\cite{XS}
\begin{eqnarray}
 \frac{d\vec{P}}{dt}=M^{\textrm{T}}\vec{P},
\end{eqnarray}
where
\begin{eqnarray}
 \vec{P}=(P_{1},P_{2},...,P_{N^{J}}),
\end{eqnarray}

\begin{eqnarray}
 \begin{cases}
      M_{ij}=\kappa_{ji}, \quad i\neq j \\
      M_{ii}=-\sum_{m}\kappa_{mi}, \quad i=j
  \end{cases}.
\end{eqnarray}
Here \emph{M} is the transition rate matrix representing the transition rates from one state (universe) to another while \emph{P}($i$) represents the fraction of comoving volume of vacuum $\Lambda_{i}$. Noticed that the Boldface represents vector. For the steady state solution ($P_{i}^{ss}$), $M^{\textmd{T}}\vec{P}^{ss}=0$, $\vec{P}^{ss}$ represents the steady state probability distribution of the vacuum states of the multiverse. We define the potential landscape as~\cite{XS,BH}
\begin{eqnarray}
\vec{U}=-\textrm{In}(\vec{P}^{ss}),
\end{eqnarray}
that is to say the potential landscape $U$ is logarithmic dependent on the steady state probability distribution or weight of the state of the multiverse $P^{SS}$ ( $\textrm{ln}(x)$ is the logarithmic function). This is analogous to the Boltzman relationship for connecting the potential energy with the equilibrium probability $e^{-\frac{U}{kT}}$ ~\cite{XS,BH}. The potential landscape $U$ thus reflects the weights of the state of the multiverse to the steady state. It tends to attract the state of the multiverse towards the lower potential or higher probability place.

As seen clearly, the transition rate matrix \emph{M} determines the evolution of the probability dynamics of the multiverse. We can decompose the driving force into the following form through the symmetrization and antisymmetrization decomposition~\cite{QZ}:
\begin{eqnarray}
 M_{ij}P_{j}^{ss}=( M_{ij}P_{j}^{ss} + M_{ji}P_{i}^{ss} )/2 +( M_{ij}P_{j}^{ss} - M_{ji}P_{i}^{ss} )/2 ,
\end{eqnarray}
define
\begin{eqnarray}
  \Delta_{ij}=\frac{1}{2 P_{j}^{ss}}(M_{ij}P_{j}^{ss}+M_{ji}P_{i}^{ss}),
\end{eqnarray}
\begin{eqnarray}
  \Theta_{ij}=\frac{1}{2 P_{j}^{ss}}(M_{ij}P_{j}^{ss}-M_{ji}P_{i}^{ss}),
\end{eqnarray}
so we have
\begin{eqnarray}
 M_{ij}= \Delta_{ij}+\Theta_{ij} .
\end{eqnarray}
Combine with the definition of the potential landscape (24), we have
\begin{eqnarray}\begin{split}
\Delta_{ij}&=\frac{1}{2} (M_{ij}+M_{ji} \frac{P_{i}^{ss}}{P_{j}^{ss}})\\
   &=\frac{1}{2} (M_{ij}+M_{ji} e^{\textrm{ln}{\frac{P_{i}^{ss}}{P_{j}^{ss}}}})\\
   &=\frac{1}{2} (M_{ij}+M_{ji} e^{U_{j}-U_{i}}).\\
\end{split}
\end{eqnarray}
Define
\begin{eqnarray}
F_{ji}^{ss}=M_{ij}P_{j}^{ss} - M_{ji}P_{i}^{ss},
\end{eqnarray}
$M_{ij}$ represents the transition rate from state \emph{j} to state \emph{i} while $P_{j}^{ss}$ represents the probability of state \emph{j}. Then $M_{ij}P_{j}^{ss}$ represents the steady state probability flux from state \emph{j} to state \emph{i}. For the same token, $M_{ji}$ represents the transition rate from state \emph{i} to state \emph{j} while $P_{i}^{ss}$ represents the probability of state \emph{i}. Then $M_{ji}P_{i}^{ss}$ represents the steady state probability flux from state \emph{i} to state \emph{j}. $F_{ji}^{ss}$  has the physical meaning of the net local steady state probability flux between \emph{i} and \emph{j}. Combine (27) and (30), we get
\begin{eqnarray}
\Theta_{ij}=\frac{F_{ji}^{ss}}{2P_{j}^{ss}}.
\end{eqnarray}
 If all the local steady state flux is zero between any \emph{i} and \emph{j} states, then there is no net input or output to or from the system. Thus the detailed balance is preserved and the system is in equilibrium state. On the other hand, if any local steady state flux is not zero, then there is a net input or output to or from the system. Thus the detailed balance is broken and the system is in nonequilibrium state.

From this decomposition, we can see that the driving force for the probability dynamics is determined by two parts. $\Delta$ is time-reversible and the detailed balance preserved part of the driving force. It is determined by the potential landscape difference or gradient. This is analogous to the usual equilibrium dynamics where the driving force is dictated by the gradient of the potential. $\Theta$ is time-irreversible. It is determined by the steady state probability flux. The steady state probability flux can be used to measure the degree of the detailed balance breaking and thus the degree of the time-irreversibility. Thus the flux quantifies the nonequilibrium (detailed balance breaking) part of the driving force of the probability evolution dynamics. If the system is in equilibrium state, $\Theta$ must be zero and equilibrium dynamics is determined by the landscape gradient alone. However, if flux is not zero, then the nonequilibrium dynamics is determined by both the landscape gradient and flux. This decomposition is in the discretized representation of the state space.

In the continuous state space, the probability evolution can be described by the Fokker-Planck equation, we can decompose the driving force into a landscape gradient and curl steady state probability flux~\cite{JW,HD,JWA}. Compared to the continuous representation, $\Delta$ is related to the landscape gradient and $\Theta$ is related to the curl flux. These two parts determine the evolution of the multiverse. While the potential landscape attracts the multiverse into certain states and provides the stability of these attractors, the flux provides a driving force to form stable flow  in the multiverse state space.

We can also decompose the driving force of the probability dynamics for the transition matrix \emph{M} into two components in another mathematical rigorous form as ~\cite{XS}:
\begin{eqnarray}
M=C+D,
\end{eqnarray}
where
\begin{eqnarray}
  C_{ij}=\begin{cases}
      max(P_{i}^{ss}\kappa_{ji}-P_{j}^{ss}\kappa_{ij},0)/P_{i}^{ss},\quad i\neq j\\
     -\sum_{m}C_{im},\quad i=j
   \end{cases},
\end{eqnarray}
\begin{eqnarray}
  D_{ij}=\begin{cases}
     min(P_{i}^{ss}\kappa_{ji},P_{j}^{ss}\kappa_{ij})/P_{i}^{ss},\quad i\neq j\\
      -\sum_{m}D_{im},\quad i=j
   \end{cases}.
\end{eqnarray}
One can easily prove that
\begin{eqnarray}
 P_{j}^{ss}D_{ij}-P_{i}^{ss}D_{ji}=0,
\end{eqnarray}
\begin{eqnarray}
P_{j}^{ss}C_{ji}-P_{i}^{ss}C_{ij}= P_{j}^{ss}\kappa_{ij}-P_{i}^{ss}\kappa_{ji}.
\end{eqnarray}
This demonstrates that we can decompose the driving force (\emph{M}) for the probability evolution into two parts. One part preserves the detailed balance (\emph{D}), this force is time-reversible,  another part breaks down the detailed balance (\emph{C}) and this part is time-irreversibility~\cite{XS}. Let us define~\cite{XS}
\begin{eqnarray}
F_{ji}= P_{j}^{ss}M_{ji}-P_{i}^{ss}M_{ij} = P_{j}^{ss}C_{ji}-P_{i}^{ss}C_{ij}
\end{eqnarray}
as the steady state flux between state $i$ and $j$.

In the next, we will show that the steady state probability flux can be further decomposed to certain cycles to form the cycle landscape of the steady state probability flux.

\subsubsection{Cycle fluxes forming flux landscapes.}

Let us look at the flux component of the driving force for the evolution of the multiverse. We can define the flux directly from the definition of the master equation
\begin{eqnarray}
\frac{dP_{i}}{dt}=\sum_{j}F_{ji},
\end{eqnarray}
while the probability flux $F_{ji}$ is defined as
\begin{eqnarray}
F_{ji}=P_{j}\kappa_{ij}-P_{i}\kappa_{ji}.
\end{eqnarray}

The master equation can be interpreted as the local conservation equation for the probability. The evolution of the probability is equal to the net flux in or out.

On the other hand, at steady state, $dP_{i}/dt=0$. If $F_{ji}=0$, the net flux is zero. This corresponds to the detailed balance and equilibrium situation. If the $F_{ji}$ is not equal zero at the steady state, then the presence of the local net flux to the system indicates that the detailed balance is broken. The steady state probability flux breaking the detailed balance becomes
\begin{eqnarray}
F_{ji}^{ss}=P_{j}^{ss}\kappa_{ij}-P_{i}^{ss}\kappa_{ji}.
\end{eqnarray}

Since both $F_{ij}^{ss}$ and $F_{ji}^{ss}$ refer to the same net local flux, we can delete this kind of redundancy. For simplicity, one can delete the one which is smaller than zero to reach the following definition~\cite{MP}:
\begin{eqnarray}
 J_{ij}=P_{j}^{ss}\kappa_{ij}-min\{P_{j}^{ss}\kappa_{ij},P_{i}^{ss}\kappa_{ji}\}.
\end{eqnarray}
One can easily prove that
\begin{eqnarray}
 J_{ij}=P_{i}^{ss}C_{ij}.
\end{eqnarray}

The definition (41) included all net local fluxes without redundancy. One can easily prove that
\begin{eqnarray}
 J_{ij}\geq0,
\end{eqnarray}
\begin{eqnarray}
 \sum_{i}J_{ij}=\sum_{i}J_{ji}.
\end{eqnarray}
Except for the null matrix, for any matrix, if it has the property (43) and (44), then one can always decompose the flux into the flux cycles (or flux loops)~\cite{XS,MP}. The procedures are outlined as follows.
Suppose $J_{i_{1}i_{2}}>0$, then from (44), one can find that $J_{i_{2}i_{3}}>0$, .... , $J_{i_{m-2}i_{m-1}}>0$, $J_{i_{m-1}i_{m}}>0$ and $i_{m}\in\{i_{1},i_{2},...,i_{m-2}\}$. Suppose that m=k, we have $J_{i_{k}i_{k+1}}>0$, $J_{i_{k+1}i_{k+2}}>0$, ..., $J_{i_{m-1}i_{k}}>0$. then $J_{i_{k}i_{k+1}}\rightarrow J_{i_{k+1}i_{k+2}}\rightarrow...\rightarrow J_{i_{m-1}i_{k}}$ represents a circulative flux, or a flux loop. Define
 \begin{eqnarray}
 \alpha_{1}=min\{J_{i_{k}i_{k+1}}, J_{i_{k+1}i_{k+2}}, ..., J_{i_{m-1}i_{k}}\}
\end{eqnarray}
 as the flux value of this loop, one can further define the circulative matrix by
\begin{eqnarray}
  r_{ij}^{(1)}=\begin{cases}
    \alpha_{1},i\in\Omega\\
    0,i\notin\Omega
   \end{cases},
\end{eqnarray}
where
\begin{eqnarray}
 \Omega=\{(i_{k},i_{k+1}),(i_{k+1},i_{k+2}),...,(i_{m-1},i_{k})\}.
\end{eqnarray}
One can easily prove that $J_{ij}^{(1)} = J_{ij}-r_{ij}^{(1)}$ still has the property (43) and (44). Therefore, one can repeat the above process and get $\alpha_{2}$, $r^{(2)}$, $J^{(2)}$, $\alpha_{3}$, $r^{(3)}$, $J^{(3)}$,... until $J^{(M+1)}=0$, where $M$ is a finite positive integer. We now have
\begin{eqnarray}
  J_{ij}=\sum_{k=1}^{M}r_{ij}^{(k)}.
\end{eqnarray}

Therefore, from the global perspective, the steady state flux can be decomposed to many cycles or loops of circulation fluxes. This forms the nonequilibrium flux landscape.

\subsection{The potential landscape in the comoving coordinate of multiverse}
\begin{figure}[tbp]
\centering
\includegraphics[width=9cm]{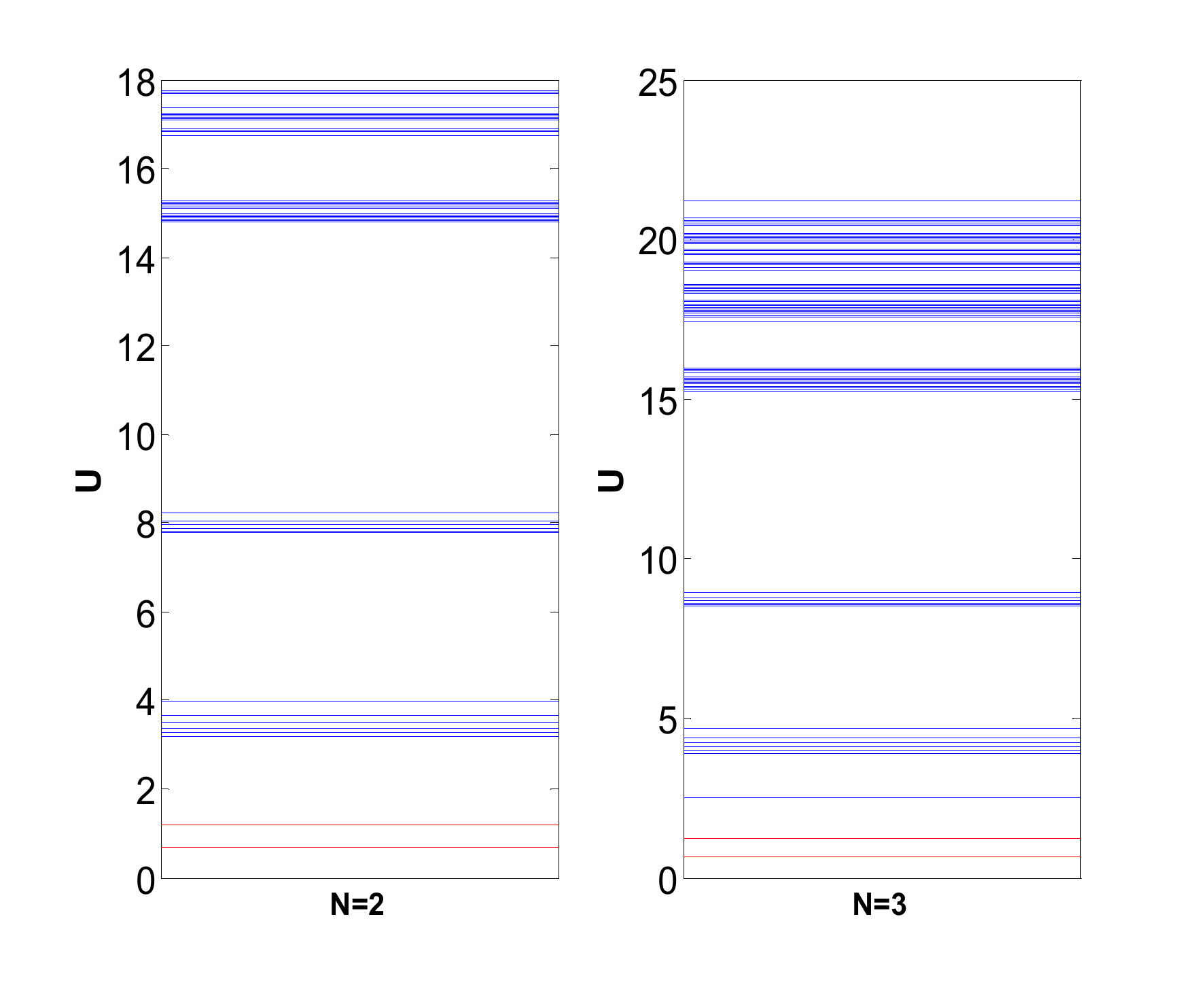}
\caption{\label{fig:2} The potential landscape spectrum. Red lines represent AdS vacuum or Minkowski vacuum, while the blue lines represent dS vacuum. The left panel includes 128 (\emph{N}=2) states (universes) and the right panel includes 2187 states (universes).}
\end{figure}

According to our definition of the potential landscape, the potential landscape spectrum is shown in figure~\ref{fig:2}, figure~\ref{fig:2} is plotted under the choice of the parameters for the string vacuum and transition among AdS, Minkowski and dS vacuum as $J=7$, $q_{1}=2.5$, $q_{2}=5.185$, $q_{3}=5.155$, $q_{4}=5.114$, $q_{5}=5.143$, $q_{6}=5.122$, $q_{7}=5.131$, $\Lambda_{bare}=-3.125$, $S_{1}=1$, $S_{2}=10^{-11}$, $S_{3}=100\times S_{2}$, $S_{4}=S_{2}$.

\begin{figure}[tbp]
\centering
\includegraphics[width=14cm]{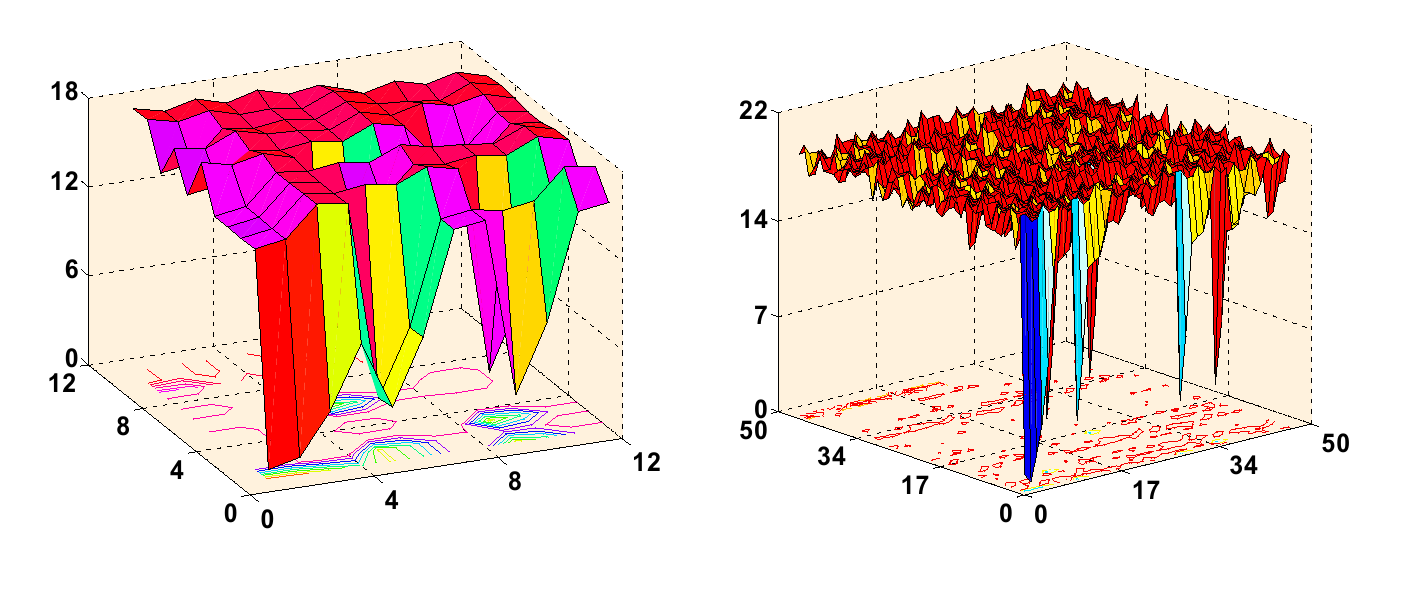}
\caption{\label{fig:3} 3-dimensional potential landscape. The left panel includes 128 (\emph{N}=2) states (universes) and the right panel include 2187 (\emph{N}=3) states (universes). The vertical axis represents the potential landscape and the colors reflect different potential levels.}
\end{figure}

The rational of the parameter choice is as follows: First, we need to specify AdS and Minkowski vacuum states, respectively. Second, the transition rates from AdS or Minkowski vacuum to the dS vacuum is assumed to be small compared with the transition rates in the opposite direction. Lastly, we assume that the transition rate from Minkowski vacuum to AdS vacuum is larger than the rate in the opposite direction.
If we arrange all these states of the multiverse on a 2 dimensional plane, the potential landscape can be shown in figure~\ref{fig:3}, The left panels of both figure~\ref{fig:2} and figure~\ref{fig:3} include 128 states ($N$=2) and the right panels of both figure~\ref{fig:2} and figure~\ref{fig:2} include 2187 states ($N$=3). The vertical axis represents the potential landscape quantifying the weight of each universe. The potential valleys of these figures represent the dominant vacuum states or universes. The color reflects the potential level. We can see that both on the left and right panels of figure~\ref{fig:2} and figure~\ref{fig:3} there are two red minimum potentials respectively with the lowest corresponding to the AdS vacuum while another corresponding to a Minkowski vacuum. The potential mimimum with the blue represents the dS vacua in figure~\ref{fig:2}. This shows the dominant vacuum as the AdS vacuum while the second dominant vacuum as the Minkowski vacuum.
We notice that there is a characteristic of the energy band (figure~\ref{fig:2}). The reason for the presence of this band is that the string theory inspired parameters $q_{2,...,7}$ are close to each other. When these string theory implied parameters become closer, the resulting characteristics of the band become clearer. Under these string parameters, the cosmological constant spectrum has the characteristic of the band. This can lead to the band structure in the potential landscape.
If there is a significant gap between the minimum and average of the potential landscape spectrum compared with the dispersion or variation, the potential landscape topography can be biased towards the dominant vacuum at the bottom. One can use a characteristic ratio for landscape topography to measure the degree of this bias~\cite{XS}, defined as
\begin{eqnarray}
 RR(U)=\frac{\delta U}{\Delta U} =\frac{<U>-U_{m}}{\sqrt{<U^{2}>-<U>^{2}}}.
\end{eqnarray}
Here $<U>$ is the average value of the potential landscape, $<U^{2}>$ is the average value of the square of the potential landscape, and $U_{m}$ is the lowest potential energy value. While the numerator represents the gap, the denominator represents the fluctuation characterized by the standard deviation.

\emph{RR} thus represents the ratio of the gap between the lowest and the average potential of the universes against the fluctuations characterized by the standard deviation through the variances. While the gap can be viewed as the slope or the bias towards the dominant vacuum, the fluctuations through the variances can be viewed as the measure of the roughness or traps of the potential landscape. A large ratio of \emph{RR} indicates a landscape with large bias towards the dominant universe against the roughness or traps. That is the landscape of the multiverse has a funneled shape towards the bottom terminal vacua (AdS and Minkowski). It also shows that the dominant state sitting at the bottom of the landscape is distinct and discriminant from the rest of the universes and therefore stable. In this sense AdS and the Minkowski vacua (universes) are usually more stable than the dS vacua (universes).

\begin{figure}[tbp]
\centering
\includegraphics[width=9cm]{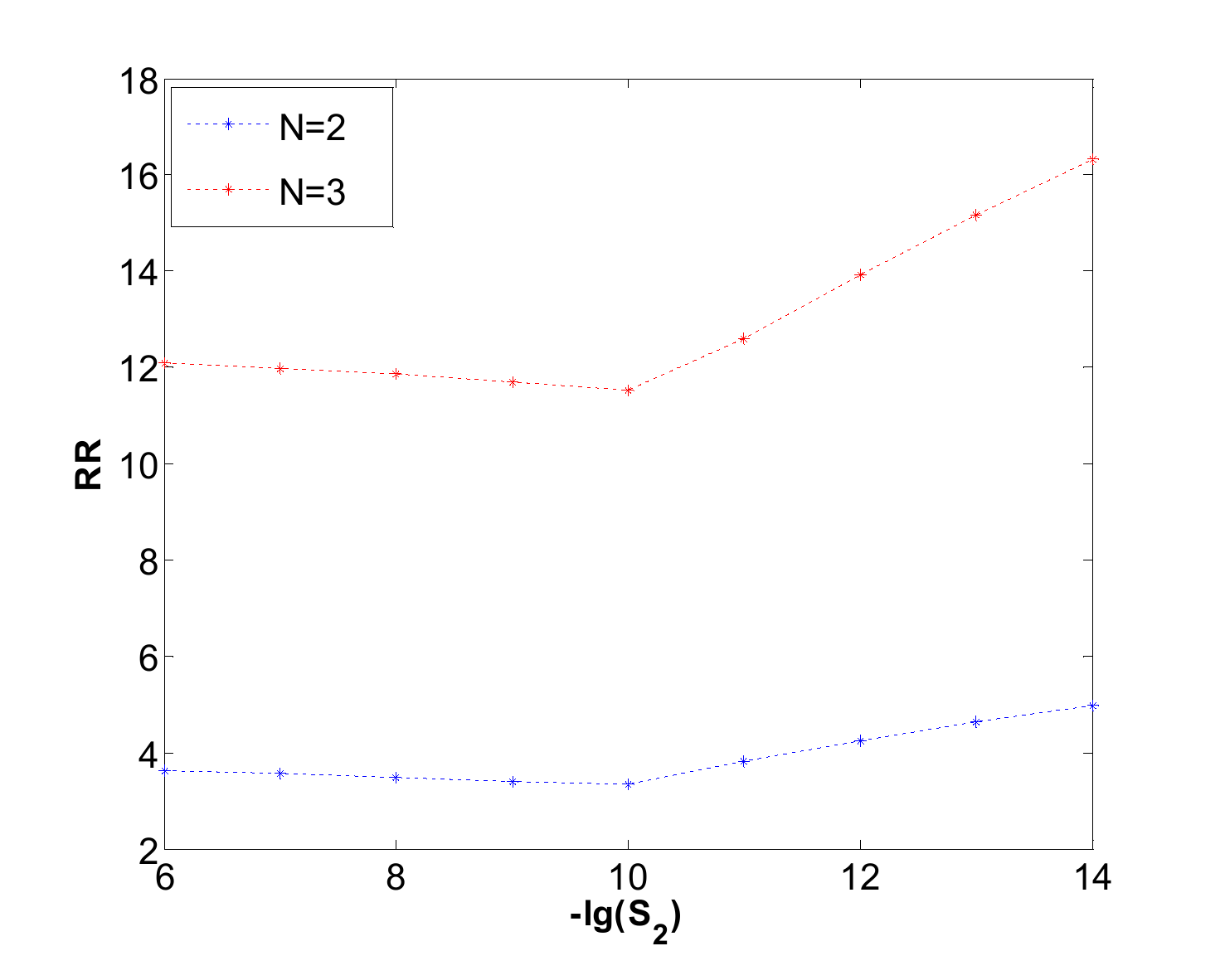}
\caption{\label{fig:4} Landscape topography ratio \emph{RR} versus the variation of the vacuum transition rates from AdS to the other universes -lg(S$_{2}$). Blue curve corresponds to \emph{N}=2 multiverse and red curve corresponds to \emph{N}=3 multiverse.}
\end{figure}

We found that when \emph{N}=2, $RR(U)=3.8166$ and when \emph{N}=3, $RR(U)=12.6087$. These two values are significantly larger than 1. This indicates that the AdS vacuum is dominant and stable against others.

Figure~\ref{fig:4} shows the \emph{RR} versus the variations of the transition rates from the AdS vacuum to the other universes, $S_{2}$. In this figure, the blue curve is for \emph{N}=2 and the red curve is for \emph{N}=3. We can see that when $S_{2}$ becomes smaller, the \emph{RR(U)} at first do not change significantly and then becomes larger. This indicates that when $S_{2}$ become smaller, the dominant vacuum becomes more distinct and discriminant from others and therefore stable. It is worthwhile to point out that our numerical simulation shows that when $S_{2}$ changes from $10^{-6}$ to $10^{-14}$ , the dominant vacuum is not always the AdS vacuum. In addition, we noticed that these two curves with \emph{N}=2 and \emph{N}=3 have similar trends against the vacuum transition rates from AdS to the other universes.

\subsection{The flux landscape in the comoving coordinate of multiverse}

\subsubsection{The flux landscape quantification}
\begin{figure}[tbp]
\centering
\includegraphics[width=10cm]{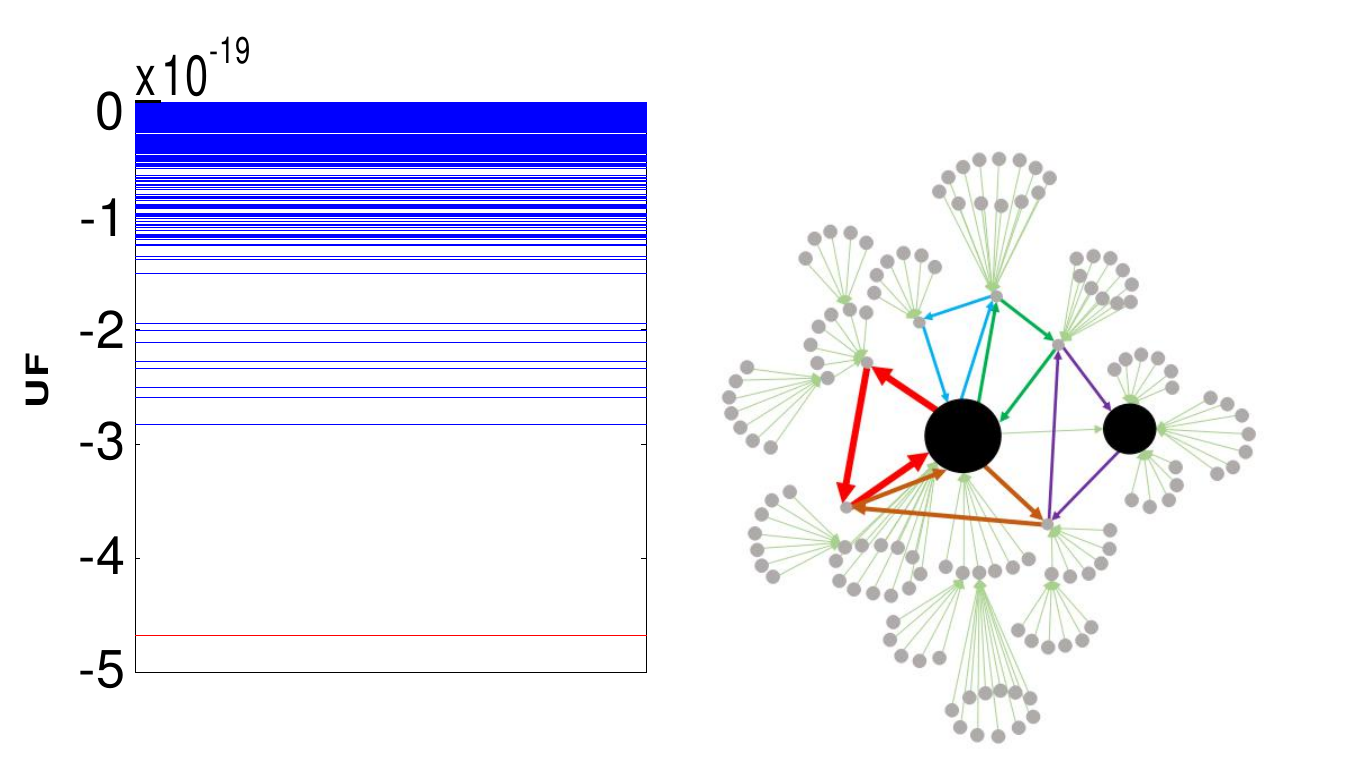}
\caption{\label{fig:5} The demonstration of the nonequilibrium flux landscape for \emph{N}=2 multiverse (128 states).  The left panel shows the flux landscape spectrum and the red lines represent the dominant cycle. The right panel is state representation of the flux landscape where each dot represents a state or universe. The flux loop with the red arrow represents the dominant loop, corresponding to the red lines in the left panel. The orange, green, blue and purple flux loops represent the second, third, fourth and fifth dominant flux loops.}
\end{figure}

The flux landscape with many flux cycles (or loops) is shown in figure~\ref{fig:5}. The parameters are set as $q_{2}=4.285$, $S_{2}$=\texttt{10$^{-17}$}, $S_{3}=2\times S_{2}$. The vertical axis of the left panel of the figure is the flux landscape value UF , which is defined as
\begin{eqnarray}
UF=-V_{f}.
\end{eqnarray}
Here, $V_{f}$ represents the value of the flux in a loop defined by (45). The red line represents the cycle with the dominant flux.
There are altogether 5073 flux loops or cycles in this multiverse connecting different universes together. The flux landscape is illustrated in the right panel of the figure~\ref{fig:5}. For the purpose of clear view, we do not present all the flux loops but only a few dominant ones. Here, different nodes represent different states or universes. The thickness of the arrows represents the magnitude of the circulation flux in the loop. The size of node represents the weight or the steady state probability of the node or the specific universe. The dominant black node represents the AdS vacuum. The second dominant black node represents the Minkowski vacuum. Other small gray nodes represent the dS vacuum. We noticed that although there are 128 states (\emph{N}=2), there are only 127 nodes. The third state does not appear in any loop. This is not strange since under these parameters,$J_{i3}=0$, for any $``i"$. Therefore, the third state cannot appear in any loop. The cycle involving the red arrows represents the dominant loop.

Since the dominant flux loop stands out from the rest of the others, it represents a limit cycle oscillation in multiverse state space. The orange arrows represent the second dominant loop and the green arrows represent the third dominant loop, while the blue and purple arrows represent the fourth and fifth dominant loops, respectively. The second, third, fourth and fifth limit cycle oscillations can also emerge but with much less chances than the dominant one.

This gives us a new angle of looking at the organization of the universes. The universes in the multiverse forms a network. Through the tunneling connections, the network of the universes is organized in a hierarchical fashion with the bottom layer or ground flux state as the dominant flux cycle connecting certain universes together. Then the hierarchical structure towers are built up by the subsequent less dominant flux cycles layer by layer or excited flux states. This new structure of the network universe indicates that the universes may emerge or function in a clustered fashion in the form of cycles.

\subsubsection{Robustness of the flux landscape}
Similar to the definition (49), we can use the definition~\cite{XS}
\begin{eqnarray}
 RR(UF)=\frac{\delta UF}{\Delta UF}=\frac{<UF>-UF_{m}}{\sqrt{<UF^{2}>-<UF>^{2}}}
\end{eqnarray}
to measure the shape of the flux landscape. $<UF>$ represents the average value of the flux landscape while $<UF^{2}>$ represents the fluctuations as the average value of the square of the flux landscape. The  $UF_{m}$ represents the smallest value of $UF$. The $RR$ is defined as the ratio of the gap between the minimum (dominant) flux potential and the average of the flux potential against the standard deviation characterizing the fluctuations in the fluxes. A high value of\emph{ RR} indicates a high discrimination of the dominant flux cycle/loop against the other ones, leading to the limit cycle oscillations among the universes in the dominant flux cycle/loop. In figure~\ref{fig:5}, $RR(UF)=11.1334$, giving arise to a distinct dominant oscillation flux cycle/loop.

Although the numerical results are different with different input parameters, the flux landscape can always emerge, as long as the detailed balance is broken and $S_{2}\neq0$. If $S_{2}=0$, this corresponds to terminal vacuum without the transitions to the other universes. Then the irreversibility comes from the time evolution and ceases to exist at long time steady state.

The flux landscape provides a new scenario and perspective for the evolution of our universe. First, due to the detailed balance breaking, the emergence of the flux gives arise to another driving force for the evolution dynamics of the multiverse. Second, the steady state flux can be decomposed to directional loops or cycles of fluxes. This generates the irreversibility and therefore the direction of time of the multiverse due to the intrinsic nonequilbrium nature of the detailed balance breaking. Note that this mechanism of time arrow is quite different from the one suggested before on the time evolution under terminal vacuum assumption where at the long time limit the evolution reaches the equilibrium preserving the detailed balance. Third, another distinct feature is that the dominant steady state probability flux cycles or loops with higher chances being seen not only involve the universes with negative or zero cosmological constant but also with the normal universes with positive cosmological constant. This indicates although the weight of our living universe may not stand out from the rest of the universes in the multiverse, it can be involved in a dominant flux cycle which increases its chance of being observed. This may help to pick up our living universe with less dependence on the anthropic principle for resolving the cosmological constant problem. Fourth, the global nature of the multiverse evolution dynamics is determined by both the underlying potential landscape and flux landscape. Fifth, the flux landscape gives arise a hierarchical structure of the organization of the universes in terms of flux cycles (or loops). Sixth, this gives a new picture of the universe evolution. From the viewpoint of the dominant cycle, the birth of our universe may be due to the transition from another universe and the death of our universe may be due to the transition to another universe. The birth and the death of the universe occur in a periodic fashion. In this sense our universe never dies. It repeats itself over and over again at a fixed amount of times by emerging from or transforming to the other universes on the dominant cycle, much like the case of a biological cell cycle~\cite{XS}. From this perspective, our universe is eternal.

\subsection{The irreversibility, the thermodynamic dissipation and time arrow of the multiverse}
\begin{figure}[tbp]
\centering
\includegraphics[width=12cm]{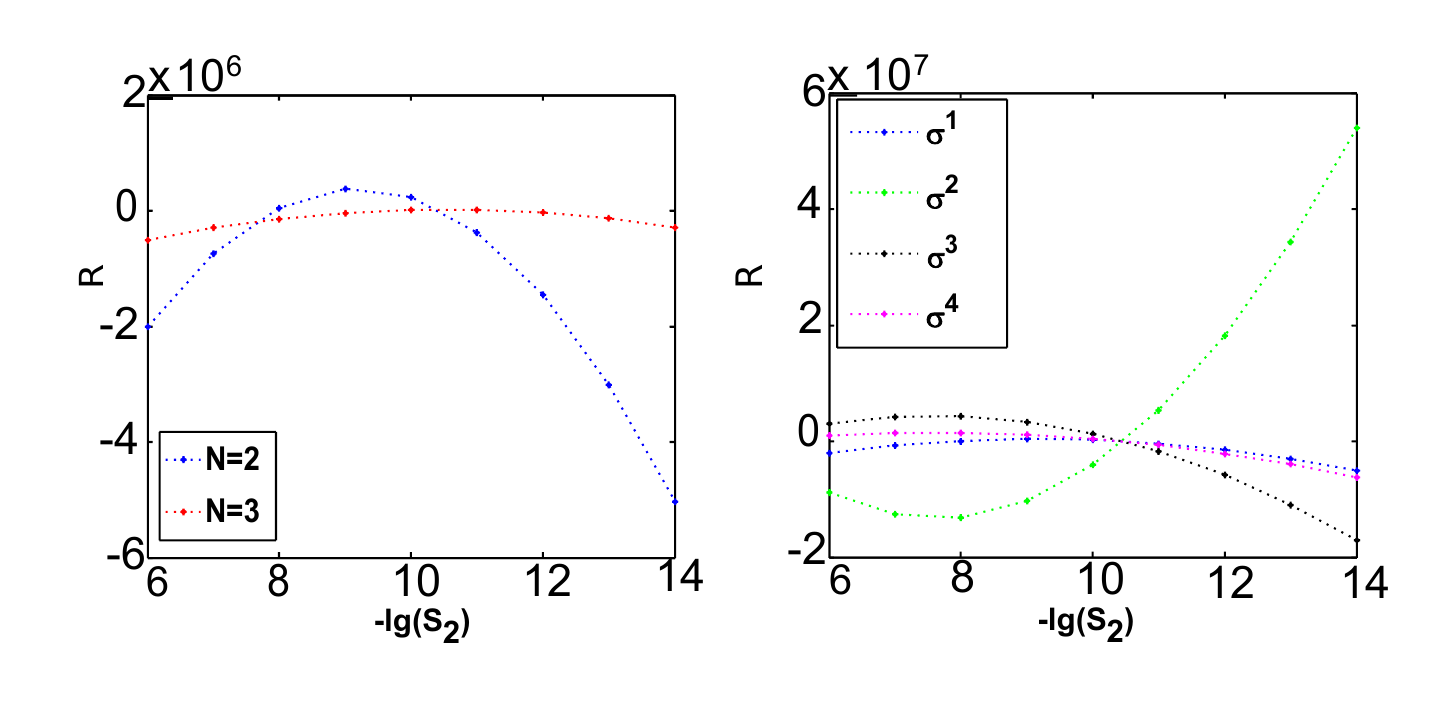}
\caption{\label{fig:6} The time irreversibility measure \emph{R} versus the variation of the vacuum transition rates from AdS to the other universes -lg(S$_{2}$). The left panel corresponds to $\sigma^{1}$ trajectory. The blue curve is for \emph{N}=2 and red curve is for \emph{N}=3 multiverses. On the right panel, there are four curves, all corresponding to \emph{N}=2 multiverse. The blue curve is for $\sigma^{1}$ trajectory, green is for $\sigma^{2}$ trajectory, black is for $\sigma^{3}$ trajectory, and purple is for $\sigma^{4}$ trajectory.}
\end{figure}

To address the time irreversibility or time arrow of the multiverse, let us study the time irreversibility of a particular trajectory. If there exists a time irreversibility on any one of the trajectories, then there is the time irreversibility of the whole system. Consider a particular trajectory specified as
\begin{eqnarray}
 \sigma=\sigma_{1}\rightarrow\sigma_{2}\rightarrow\sigma_{3}\rightarrow...\rightarrow\sigma_{n}.
\end{eqnarray}
Here  $\sigma_{m}$ ( m=1,2,3,... ) represents different vacuum state. $\sigma$ represents a transition trajectory at first from the state $\sigma_{1}$  to $\sigma_{2}$, then from  $\sigma_{2}$ to $\sigma_{3}$ ,..., and finally from  $\sigma_{n-1}$ to $\sigma_{n}$.
We consider the following function ~\cite{RJ}
\begin{eqnarray}
  R=\texttt{ln}[\frac{\kappa_{\sigma_{n},\sigma_{n-1}}...\kappa_{\sigma_{3},\sigma_{2}}\kappa_{\sigma_{2},\sigma_{1}}P(\sigma_{1})}{\kappa_{\sigma_{1},\sigma_{2}}\kappa_{\sigma_{2},\sigma_{3}}...\kappa_{\sigma_{n-1},\sigma_{n}}P(\sigma_{n})}]
\end{eqnarray}
to measure the time irreversibility of the trajectory (52). Here $P(\sigma_{n})$  represents the fraction of the vacuum state $\sigma_{n}$.  This gives rise to the ratio of the forward in time transition probability against the backward in time transition probability. An equal forward and backward transition probability will lead to an equilibrium with a zero value in \emph{R}. This indicates the time reversibility. The magnitude of \emph{R} provides a measure of how forward rates are different from the backward rates in time. Therefore, \emph{R} is a measure of the time irreversibility or time arrow. Figure ~\ref{fig:6} shows the value of \emph{R} or time irreversibility versus the variation of the parameter  $S_{2}$ for some trajectories
\begin{eqnarray}\begin{split}
  &\sigma^{1}=\sigma_{1}\rightarrow\sigma_{2}\rightarrow...\rightarrow\sigma_{9}\rightarrow\sigma_{10},\\
  &\sigma^{2}=\sigma_{1}\rightarrow\sigma_{2}\rightarrow...\rightarrow\sigma_{10}\rightarrow\sigma_{11},\\
  &\sigma^{3}=\sigma_{1}\rightarrow\sigma_{2}\rightarrow...\rightarrow\sigma_{11}\rightarrow\sigma_{12},\\
  &\sigma^{4}=\sigma_{1}\rightarrow\sigma_{2}\rightarrow...\rightarrow\sigma_{12}\rightarrow\sigma_{13}.
\end{split}
\end{eqnarray}
Noticed that we use $\sigma^{i}$ (\emph{i}=1,2,3,... ) to represent different transition trajectories and use $\sigma_{j}$ (\emph{j}=1,2,3,... ) to represent different vacuum states.

The left panel of figure~\ref{fig:6} is plotted based on the trajectory $\sigma^{1}$. The blue curve is for \emph{N}=2 and the red curve is for \emph{N}=3. The left panel shows that the appearance of the irreversibility or time arrow since in most of the parameter range \emph{R} is significantly different from zero. The right panel of figure~\ref{fig:6} is plotted based on the trajectories $\sigma^{1,2,3,4}$ and \emph{N}=2. The right panel of figure~\ref{fig:6} shows that \emph{R} can qualitatively be used to characterize the irreversibility of the system. In order to quantitatively describe the irreversibility and link to the thermodynamics,we consider the average value  $<R>$ of the \emph{R} ~\cite{ME}. We noticed that \emph{R} is a function of the trajectory $\sigma$, different trajectory $\sigma$  gives rise to different value of \emph{R}. Therefore,  the average value $<R>$  is
\begin{eqnarray}
 <R>=\sum_{\sigma}P\{\sigma\}R\{\sigma\}.
\end{eqnarray}
Here, the summation is over all possible trajectories,  $P(\sigma)$   is the weight of the trajectory $\sigma$ , for a trajectory defined by (52),
\begin{eqnarray}
 P(\sigma)=\kappa_{\sigma_{n},\sigma_{n-1}}....\kappa_{\sigma_{3},\sigma_{2}}\kappa_{\sigma_{2},\sigma_{1}}P(\sigma_{1}).
\end{eqnarray}
All the trajectories can be classified by including two vacuum states, three vacuum states, four vacuum states,...,n vacuum states,... Therefore, equation (55) can be written as
\begin{eqnarray}\begin{split}
  <R>=\sum_{ij}&P_{i}\kappa_{ji}\texttt{ln}\frac{P_{i}\kappa_{ji}}{P_{j}\kappa_{ij}}+\sum_{ijk}P_{i}\kappa_{ji}\kappa_{kj}\texttt{ln}\frac{P_{i}\kappa_{ji}\kappa_{kj}}{P_{k}\kappa_{jk}\kappa_{ij}}\\
      &+\sum_{ijkl}P_{i}\kappa_{ji}\kappa_{kj}\kappa_{lk}\texttt{ln}\frac{P_{i}\kappa_{ji}\kappa_{kj}\kappa_{lk}}{P_{l}\kappa_{kl}\kappa_{jk}\kappa_{ij}}+...
 \end{split}
\end{eqnarray}
On the right hand side of the formula (57), the first term means the sum over all trajectories including two vacuum states. The second term means the sum over all trajectories including three vacuum states,...Define:
\begin{align}
&EPR=\sum_{ij}P_{i}\kappa_{ji}\texttt{ln}\frac{P_{i}\kappa_{ji}}{P_{j}\kappa_{ij}},\\
&\Delta_{1}=\sum_{ijk}P_{i}\kappa_{ji}\kappa_{kj}\texttt{ln}\frac{P_{i}\kappa_{ji}\kappa_{kj}}{P_{k}\kappa_{jk}\kappa_{ij}},\\
&\Delta_{2}=\sum_{ijkl}P_{i}\kappa_{ji}\kappa_{kj}\kappa_{lk}\texttt{ln}\frac{P_{i}\kappa_{ji}\kappa_{kj}\kappa_{lk}}{P_{l}\kappa_{kl}\kappa_{jk}\kappa_{ij}},
\end{align}
then
\begin{eqnarray}
 <R>=EPR+\Delta_{1}+\Delta_{2}+...
\end{eqnarray}
\emph{EPR} is the entropy production rate~\cite{JS}. the precise relation between \emph{EPR} and the irreversibility revealed by the famous fluctuation theorem~\cite{GE}.We noted that
\begin{eqnarray}\begin{split}
EPR&=\sum_{ij}P_{i}\kappa_{ji}\texttt{ln}\frac{P_{i}\kappa_{ji}}{P_{j}\kappa_{ij}}\\
   &=\frac{1}{2}\sum_{ij}(P_{i}\kappa_{ji}-P_{j}\kappa_{ij})\texttt{ln}\frac{P_{i}\kappa_{ji}}{P_{j}\kappa_{ij}}\\
   &=\frac{1}{2}\sum_{ij}F_{ij}^{ss}\texttt{ln}\frac{P_{i}\kappa_{ji}}{P_{j}\kappa_{ij}}.\\
\end{split}
\end{eqnarray}
This is the connection between the \emph{EPR} and the steady state probability flux. The steady state flux gives the dynamical origin of the nonequilibriumness through the detailed balance breaking, while the steady state \emph{EPR} provides the thermodynamic origin of the nonequilibirumness for maintaining the steady state.

We can define that
\begin{align}
&<R>_{1}= EPR,\\
&<R>_{2}= EPR+\Delta_{1},\\
&<R>_{3}= EPR+\Delta_{1}+\Delta_{2}.
\end{align}
For the potential landscape in figure~\ref{fig:2}, when \emph{N}=2, $<R>_{1}=\texttt{2.3747}\times\texttt{10}^{-\texttt{9}}$; when \emph{N}=3, $<R>_{1}=\texttt{1.4419}\times\texttt{10}^{-\texttt{8}}$ . Again, the entropy production rate is not zero indicating the nonequilibrium thermodynamic cost is needed to maintain the irreversibility and time arrow.

\begin{figure}[tbp]
\centering
\includegraphics[width=9cm]{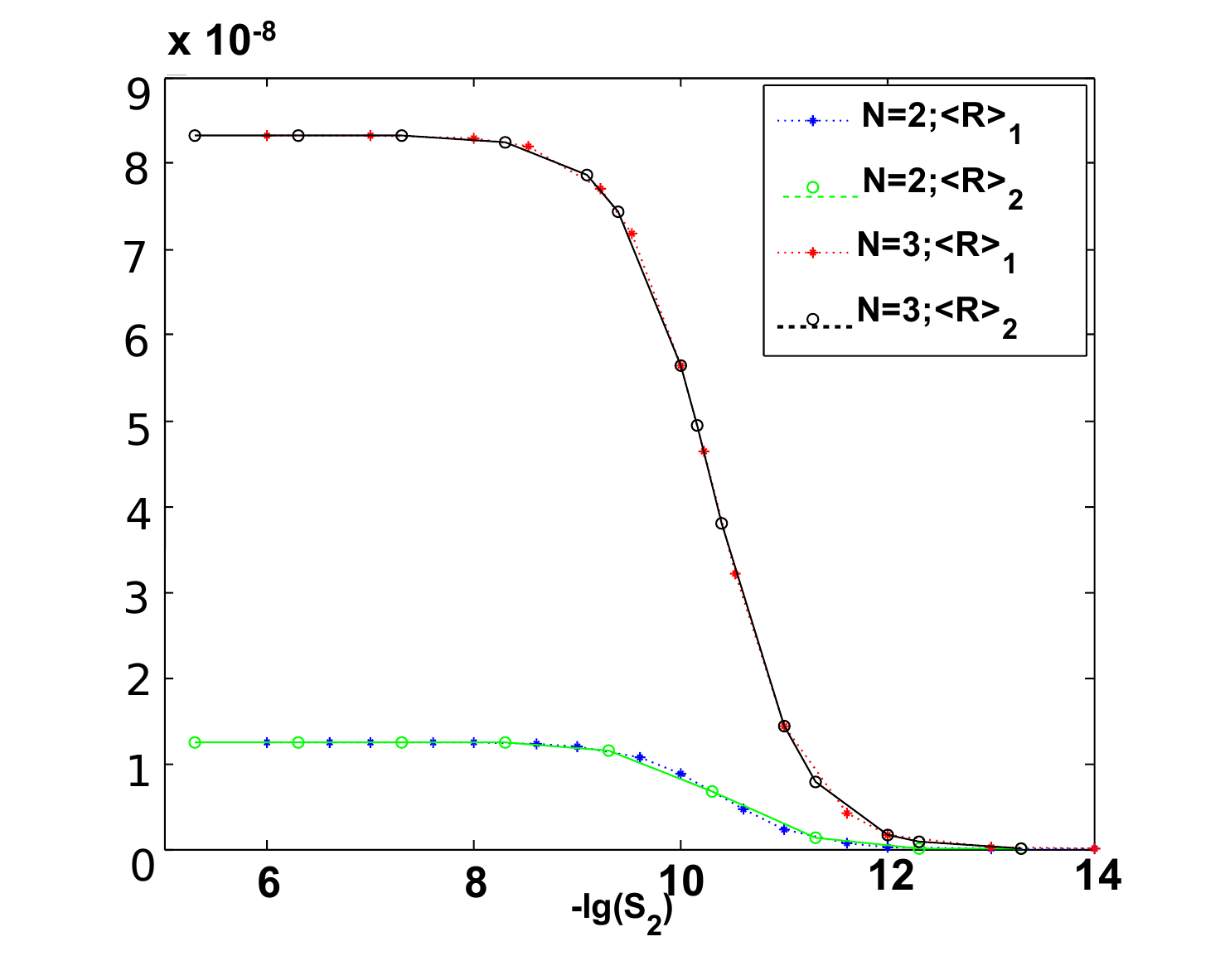}
\caption{\label{fig:7} The time irreversibility measures $<R>_{1,2}$ versus the variation of the vacuum transition rates from AdS to the other universes -lg(S$_{2}$). The blue curve is $<R>_{1}$ versus the variation of -lg(S$_{2}$) for \emph{N}=2 multiverse. The green curve is $<R>_{2}$ versus the variation of -lg(S$_{2}$) for \emph{N}=2 multiverse. The red curve is $<R>_{1}$ versus the variation of -lg(S$_{2}$) for \emph{N}=3 multiverse. The black curve is $<R>_{2}$ versus the variation of -lg(S$_{2}$) for \emph{N}=3 multiverse.}
\end{figure}

Figure~\ref{fig:7} shows the first (average) and second order (fluctuation) of $<R>$ versus the variation of the parameter $S_{2}$. Larger transition rate from AdS to other universes will give arise to more degree of detailed balance breaking and therefore the time arrow of the multiverse. The blue dotted curve and the green dash-dotted curve represent the $<R>_{1}$ and $<R>_{2}$ respectively for \emph{N}=2. Similar two curves representing the $<R>_{1}$ and $<R>_{2}$ respectively are plotted for \emph{N}=3.

Figure~\ref{fig:7} shows that the following approximation is a good one:
\begin{eqnarray}
<R>\approx EPR.
\end{eqnarray}
Therefore, we can use both $<R>$ and \emph{EPR} to describe the irreversibility of the system. In other words, $<R>$ and \emph{EPR} can provide the thermodynamic measure of the time arrow of the multiverse in comoving coordinate. Here, the origin of the time arrow is the detail-balance breaking. This is different with the time arrow in~\cite{RBO} which comes from the dynamic evolution of our universe. We should point out that when $S_{2}$ is strictly equal to zero, the AdS vacuum and the Minkowski vacuum become the terminal vacuum. They cannot decay to another vacuum. In this case, the formula (63) or (64) cannot be directly used to describe the arrow of time. Because in this case, the fraction of the AdS vacuum is 1 and others are zero. The formula (63) and (64) are ill-defined.

\section{CONCLUSIONS}

In this work, we studied the nonequilibrium evolution of the multiverse in the comoving coordinate in detail. We uncovered that the driving force of the multiverse evolution in general is determined by the global landscape quantified by the steady state probability distribution and the steady state probability flux. While the landscape attracts the multiverse to the steady state basins, the flux gives arise to cycles and loops associating certain universes together. The emergence of the flux loops provides a signature and a quantitative measure of the degree of detailed balance breaking-nonequilibriumness. In other words, the landscape of the multiverse is uncovered and quantified in this work.  However, the multiverse evolution dynamics is driven by not only the landscape but also the flux. This is very different from the conventional picture of the dynamics determined by the landscape gradient for equilibrium systems under detailed balance.

The vacuum of the universes with positive cosmological constant can be transformed from each other through tunneling. Recent studies show the possibilities of the bounce of contracting universe with negative cosmological constant back to the expanding universe with positive cosmological constant~\cite{EF,EFA,WF,WFI,AL}. This can lead to the breaking of the detailed balance~\cite{JG,FD,SFVF}. The flux associated with the detailed balance breaking provides the dynamical origin of the irreversibility and time arrow of the multiverse. On the other hand, the entropy production is associated with the flux. Thus, it provides the thermodynamic origin or cost for maintaining the irreversibility and time arrow of the multiverse. In contrast to the time evolution argument of the underlying detailed balance system with terminal vacuum for generating the irreversibility, the current approach emphasizes the possibility of the time arrow generated by the detailed balance breaking even at long time steady state. This provides a new mechanism for the time arrow for the multiverse.

Furthermore, the potential landscape shows a funneled shape dominated by the contracting universes with negative cosmological constant and flat universes with zero cosmological constant. On the other hand, the flux loops form a flux landscape. The flux landscape also shows a funneled shape dominated by certain flux cycles or loops. The dominant flux loops or cycles can involve not only the contracting and flat universes, but also expanding universes with positive cosmological constant. The dominant loop gives arise to the associations of certain different kinds of the universes together and this leads to the oscillation cycles among these universes in the multiverse evolution. In other words, a universe on the dominant cycle can appear due to the transition from the other universes on the same cycle and can also disappear due to the transition to the other universes on the same cycle. This occurs periodically. If our universe is on this dominant cycle, then it can be born due to the transition from another universe and die due to the transition to another universe in a periodic manner. In this sense, our universe never dies. The birth and death of our universe go through cycles. Therefore, the universes on the dominant cycle oscillate from one universe to another and repeatedly appear and disappear under a fixed period of times.

Moreover, although our living universe with small positive cosmological constant does not necessarily have significantly higher probability compared to others in the multiverse, it can lie in the dominant flux cycles. If that is the case, then the chance of being observed can be significantly enhanced. Thus, this may provide a boost in addition to the anthropic principle for selecting our living universe.

The presence of the steady state probability flux breaks the detailed balance. This leads to the multiverse as an intrinsic nonequilibirum system. Even in steady state, the equilibrium is not reached since the detailed balance is not preserved and there is a net flux. In other words, the time arrow of the multiverse originated from this intrinsic nonequilibriumness exists at all times (even at the long time limit). This is in contrast to the case where the time arrow is generated during the evolution of the multiverse and ceases to exist at the long time limit.

\section*{Acknowledgments}
 Hong Wang thanks for support from Natural Science Foundation of China No.21721003.

\end{document}